\documentclass[%
 reprint,
superscriptaddress,
prb,
floatfix,
]{revtex4-2}

\usepackage{graphicx}
\usepackage{dcolumn}
\usepackage{bm}
\usepackage{natbib}
\usepackage{ulem}
\usepackage{xcolor}
\usepackage{amsmath} 
\usepackage{amssymb}
\usepackage{hyperref}
\usepackage{multirow}

\begin{document}

\preprint{APS/123-QED}

\title{Harnessing magnetic anisotropy for nonlinear magnetization precession and spin waves}

 \author{P. I. Gerevenkov}
 \email{petr.gerevenkov@mail.ioffe.ru}
 \homepage{http://www.ioffe.ru/ferrolab/}
 	\affiliation{Ioffe Institute, 194021 St. Petersburg, Russia}
 \author{L. A. Shelukhin}
 	\affiliation{Ioffe Institute, 194021 St. Petersburg, Russia}
 \author{Ia. A. Filatov}
 	\affiliation{Ioffe Institute, 194021 St. Petersburg, Russia}
 \author{P. A. Dvortsova}
 	\affiliation{Ioffe Institute, 194021 St. Petersburg, Russia}
 \author{A. M. Kalashnikova}
 	\affiliation{Ioffe Institute, 194021 St. Petersburg, Russia}



\begin{abstract}
	The nonlinearity of magnetization precession and spin waves is a cornerstone of contemporary magnonics.
	We investigate nonlinear magnetization dynamics in a thin epitaxial iron film driven by femtosecond laser pulses in regimes of homogeneous precession and propagating magnetostatic spin wave packets.
	The magnetization precession anharmonicity, the generation of higher-order harmonics, and the dynamical rectification are experimentally demonstrated.
	The numerical solution of the non-linearized Landau-Lifshitz-Gilbert equation reveals that these effects stem from the asymmetry in the energy potential and are essentially thresholdless.
	This asymmetry is readily achievable when an external magnetic field with a strength comparable to the magnetic anisotropy field is applied close to the hard axis.
	This work establishes a connection between the geometry of the energy profile and nonlinear responses, paving the way for designing magnonic devices with controlled harmonic generation and nonlinear spin wave interaction.

\end{abstract}

\maketitle


\section{\label{sec:intro} Introduction}

In wave physics, nonlinearity is traditionally understood as a wave-induced change in a medium’s parameters (e.g. dispersion, damping) that enables wave interactions~\cite{breitbach2024nonlinear}.
Such interactions in magnetic systems are essential for applications ranging from all-magnonic transistors~\cite{chumak2014magnon,ge2024nanoscaled} to computing~\cite{Wang2020}, neuromorphic devices~\cite{Papp2021,lutsenko2026nonlinear,breitbach2025all}, and frequency-comb generation~\cite{wang2024enhancement,wang2021magnonic,khivintsev2011nonlinear,guo2026stimulated}.
Within the macrospin approximation (setting aside processes like the splitting of one oscillation into several~\cite{Dreyer2022,demidov2011nonlinear,demokritov2006bose}), research has largely focused on the type-I Suhl nonlinearity, where the demagnetizing fields change due to the reduction of the longitudinal magnetization component during large-angle precession~\cite{suhl_1957_instability, Suhl_NonLinFer, livesey2015nonlinear, guo2015nonlinear,flebus20242024}.
The onset of nonlinearity is typically associated with high-amplitude precession or waves, while the underlying physical mechanisms are often not detailed.
Conventionally, the nonlinear regime is identified by a resonant frequency shift~\cite{Khivintsev_PRB2010, wismayer2012nonlinear, guo2015nonlinear,edwards2012parametric} or the generation of higher harmonics~\cite{zhang2023generation, marsh2012nonlinearly,he2014probing,cheng2013high,hudl2019nonlinear}.
However, at large amplitudes and strong precession ellipticity, a purely geometric second harmonic can appear~\cite{solovev2021second, nikolaev2025resonant,nikolaev2024resonant}. 
In that case, only the double frequency is generated, and waves may not interact if the dispersion remains amplitude-independent~\cite{solovev2022second}.

In this Article, we examine a general mechanism that exploits the asymmetric energy potential created by a magnetic field and magnetic anisotropy, leading directly to nonlinear magnetization precession.  
This mechanism enables experimental demonstration of higher harmonics up to the 4th order, magnetization precession rectification, and a nonlinear shift of the resonance frequency for both uniform precession and spin wave packets in a thin Fe film. 
The proposed mechanism is not reducible to type-I Suhl nonlinearity and occurs for magnetization deviations below one degree.

\section{\label{sec:experimental} Experimental And Simulations Details}

We employed time-resolved magneto-optical Kerr effect to study magnetization dynamics in a thin Fe film.
Experiments were performed on an epitaxial Fe~(001) film with thickness $d = 20$\,nm grown by pulsed laser deposition on a MgO~(001) substrate~\cite{suturin2022laser,dvortsova2022technological}.
The film exhibits cubic magnetocrystalline anisotropy with easy axes along $\langle$110$\rangle$ MgO directions and strain-induced anisotropy with an (001) easy-plane~\cite{khodadadi2020conductivitylike,tournerie2008plane,solano2022spin}.
The external magnetic field ($\mathbf{H}_{ext}$) lies in the film plane and makes an angle $\phi_H$ with the hard axis (HA).
Ultrafast heating by a 190-fs laser pulse rapidly reduces the effective anisotropy field via demagnetization and a change of the magnetic anisotropy parameter~\cite{kalashnikova2023ultrafast}.
As a result, magnetization precession is triggered within the pumped volume, and its frequency is controlled by the thermally altered anisotropy field~\cite{carpene2010ultrafast,gerevenkov2021effect}.
A magnetostatic wavepacket is also launched and propagates laterally away from the laser spot according to the dispersion governed by the equilibrium anisotropy and $\mathbf{H}_{ext}$~\cite{Khokhlov_PRAppl2019,filatov2022spectrum}.
To monitor the laser-induced dynamics, time- and spatially resolved pump-probe measurements were performed as described elsewhere~\cite{Khokhlov_PRAppl2019} and in Sec.~I of Suppl. Material~\cite{Supplemental}.

To analyze the precession and spin waves, we numerically solved the non-linearized Landau-Lifshitz-Gilbert (LLG) equation in the time domain (Sec.~II of Suppl. Material~\cite{Supplemental}) and performed micromagnetic simulations with mumax3~\cite{vansteenkiste2014design} (see~\cite{gerevenkov2023unidirectional} and Sec.~III of Suppl. Material~\cite{Supplemental}), including the time-dependent and spatially localized change of magnetization saturation $M_S$ and cubic anisotropy parameter $K_C$.

The magnetic parameters of the film under study are determined in Ref.~\cite{gerevenkov2026role}.
The parameter values before excitation: $M_S = 1.72$\,MA~m$^{-1}$, $K_C = 41.8$\,kJ~m$^{-3}$, and uniaxial anisotropy parameter $K_U = 102.2$\,kJ~m$^{-3}$.
Importantly, the Gilbert damping is $\alpha_G = 4 \cdot 10^{-3}$, comparable to the lowest reported for epitaxial Fe films at room temperature~\cite{khodadadi2020conductivitylike} and essential for reliable demonstration of nonlinearities in both magnetization precession and spin waves.
Following Ref.~\cite{gerevenkov2021effect}, laser excitation is modeled as an instantaneous temperature increase $\Delta T = 241$~K (for average fluence $F = 14$~mJ~cm$^{-2}$) followed by exponential recovery with a characteristic time of 0.74\,ns.
The thermal dependence of $M_S$ is reconstructed from experiment~\cite{crangle1971magnetization}, and $K_C (T)$ is obtained using the power law $\left(K_C(T)/K_C(0)\right) = \left(M_S(T)/M_S(0)\right)^{10}$~\cite{zener1954classical,akulov1936quantentheorie}, where the relative change in $K_C$ scales as the 10th power of the relative change in $M_S$.

\section{\label{sec:results_discussion} Results and Discussion}

\subsection{\label{sec:precession} Magnetization precession}

\begin{figure}[ht]
	\includegraphics[width= 0.9\linewidth]{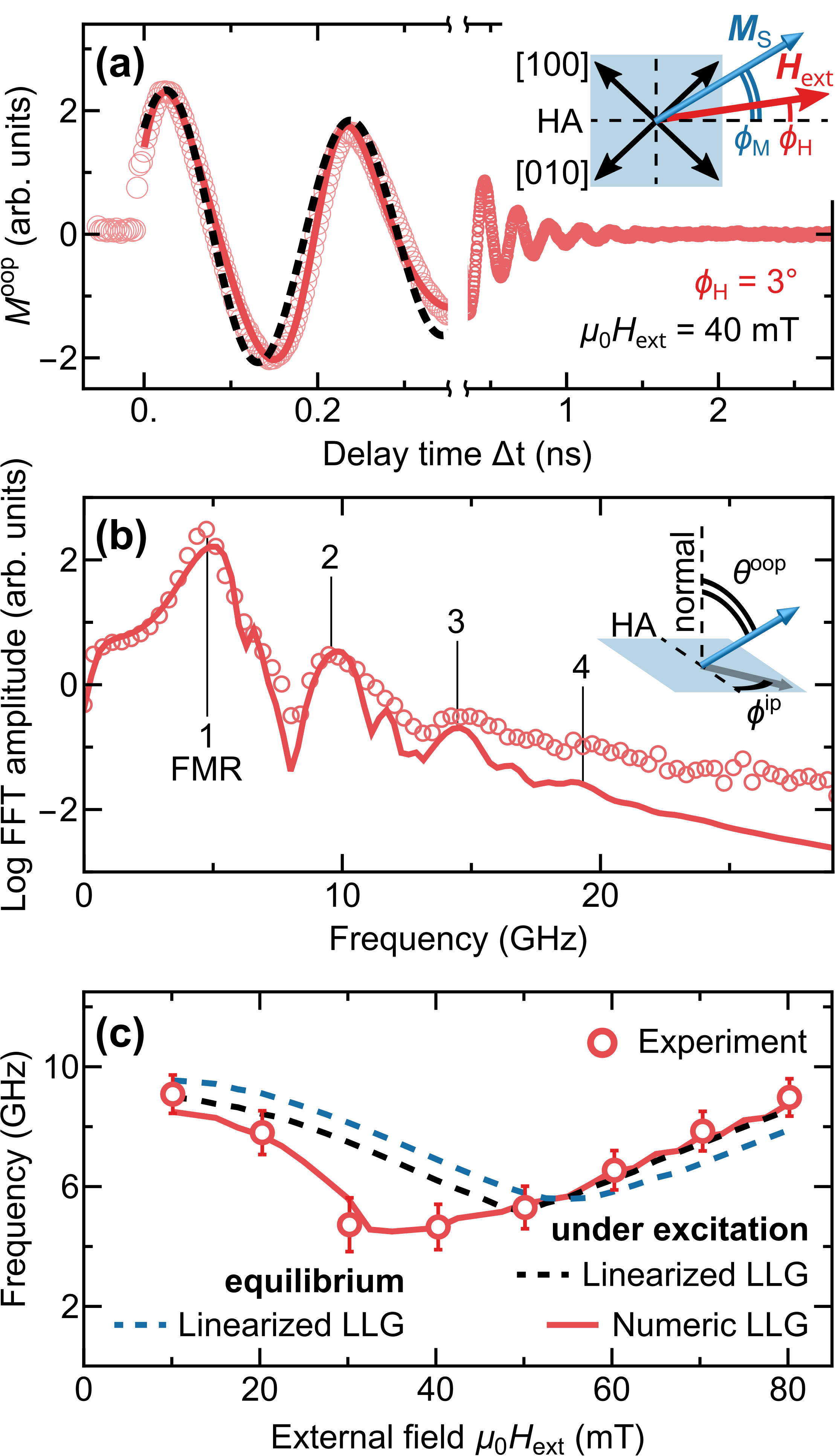}
	\caption{\label{fig:high_harmonics}
		{\bf Anharmonicity and higher harmonics generation.}
		(a) Experimental pump-probe signals (points), proportional to the out-of-plane magnetization component $M^{oop}$. 
		Measurements were performed at $\mu_0 H_{ext} = 40$\,mT, directed at $\phi_H=3\,^\circ$. 
		The solid red and dashed black lines show the numerical LLG solution and a single-frequency damped sine, respectively.
		The insert in (a) sketches the in-plane anisotropy axes and angle definitions: easy axes (double-headed arrows), hard axes (dashed lines).
		(b) FFT of the experimental signal and the numerical LLG solution from (a).
		Harmonics are numbered (1 is the eigen frequency).
		The inset in (b) shows a coordinate system.
		(c) Field dependencies of the experimental eigen frequency (points), numeric (solid red line) and linearized under excitation (dashed black line) and equilibrium (dashed blue line) LLG solutions.
		Error bars show the standard deviation of the experimental FFT peaks.
	}
\end{figure}

To explore the nonlinear precession and spin waves, we rely on the interplay between magnetic anisotropy and $\mathbf{H}_{ext}$, which shapes the magnetic energy profile (the free energy as a function of the in-plane magnetization angle).
When a magnetic field is applied close to the HA with a magnitude near the effective anisotropy field, the free energy profile becomes intrinsically asymmetric and nonparabolic, which favors nonlinearity.
Therefore, we design the experiment such that this condition is met after the laser pulse reduces $M_S$ and $K_C$.
We apply $\mu_0 H_{ext} = 40$\,mT at an angle $\phi_H = 3$\,$^{\circ}$ to the HA [inset in Fig.~\ref{fig:high_harmonics}\,(a)].
Using the laser‑altered magnetic parameters, we compute the magnetization dynamics within the macrospin approximation.
The resulting out‑of‑plane magnetization component is shown by the red line in Fig.~\ref{fig:high_harmonics}\,(a).
It clearly deviates from a damped sinusoidal function (black dashed line in the same panel), indicating precession anharmonicity and signifying its nonlinear origin~\cite{gaponov1979li}.
The FFT spectrum [red line in Fig.~\ref{fig:high_harmonics}\,(b)] exhibits both the fundamental precession mode (marked as FMR) and its higher harmonics up to fourth order.

To confirm that such pronounced nonlinearity can be observed experimentally, we plot in Fig.~\ref{fig:high_harmonics}\,(a, b) the signal obtained with weakly-focused, spatially-overlapped pump and probe pulses~\cite{Supplemental} and its FFT.
The laser fluence was chosen to match the reduction of magnetic parameters used in the calculations.
There is a striking agreement between the calculated and experimental curves, which exhibit the same features.
Note that the equidistant peaks in the FFT, together with the fact that the spectrum is reproduced by the macrospin approximation and the excitation is nearly uniform across the thickness, rule out thickness spin-wave resonances (which produce a quadratically spaced spectrum) as the source of anharmonicity~\cite{weber1968SWR, seavey1958direct}.
The 1st-order spin-wave resonance for a 20-nm-thick Fe film is about 20\,GHz~\cite{gladii2017spin}, comparable only to the fourth harmonic in our experiment.

To reveal under which conditions nonlinearity is most pronounced, we plot in Fig.~\ref{fig:high_harmonics}\,(c) the dependence of the FMR frequency on the external magnetic field obtained from experiment (symbols) and from numerical calculations (red solid line).
The two are in good agreement.
In contrast, near the critical field of the laser-excited film (40~mT), the solution of the linearized LLG equation~\cite{smit1955ferromagnetic} (black dashed line) does not match the experimental results.
Higher harmonics also appear in the same field region.
Importantly, nonlinearity causes qualitative changes in the field  dependence of the FMR frequency that can not be explained by the laser-induced reduction of $M_S$ and $K_C$ alone.
To emphasise this, we plot in Fig.~\ref{fig:high_harmonics}\,(c) the linearized LLG solution for the equilibrium parameters of the film (dashed blue line).
Clearly, disregarding nonlinearity can obscure the conclusions about the laser-induced changes of the magnetic state.

\subsection{\label{sec:theory} Theory}

\begin{figure}[ht!]
	\includegraphics[width= 0.8\linewidth]{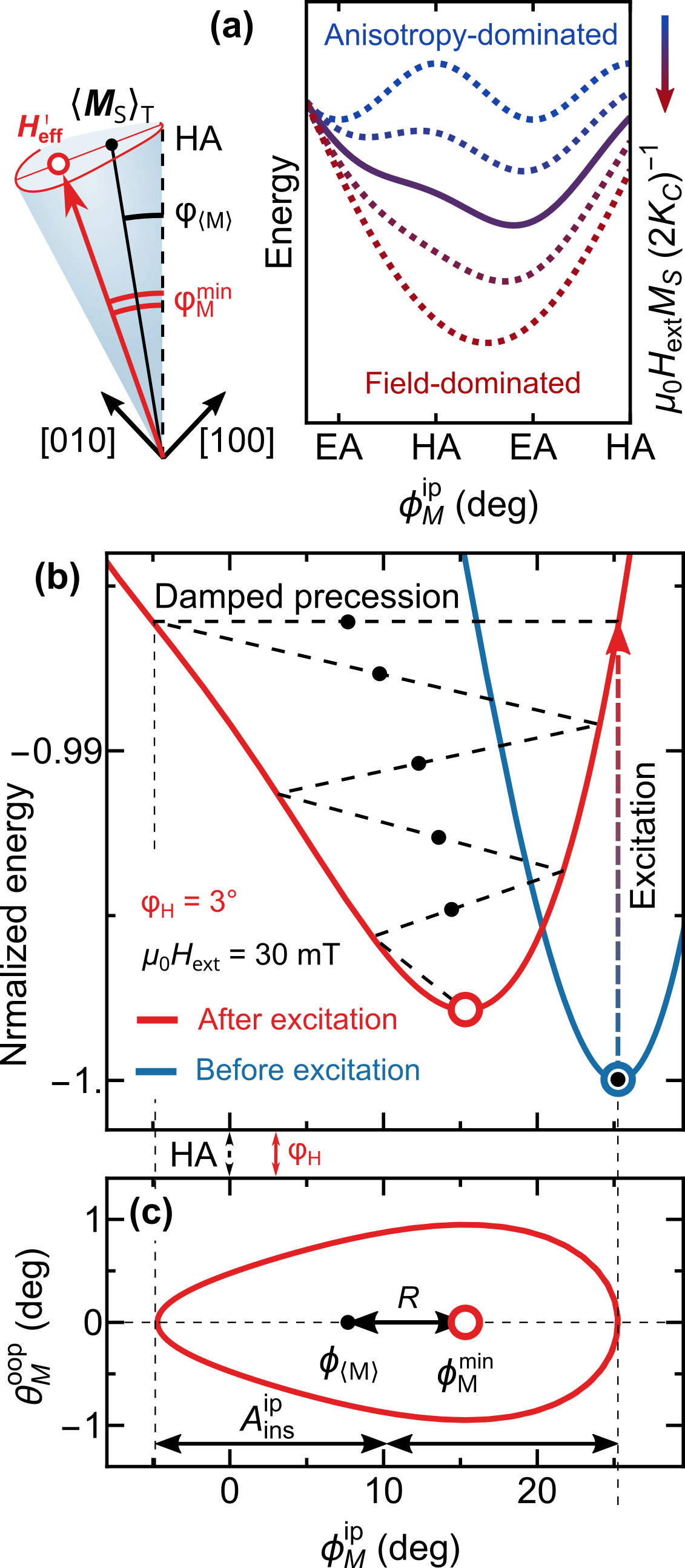}
	\caption{\label{fig:energy_profile}
		{\bf Energy profile asymmetry as a source of anharmonicity.}
		(a) Schematic diagram of the effective field direction and the period-averaged magnetization relative to the HA (left).
		Illustration of how competing contributions from the external field and in-plane cubic anisotropy create potential asymmetry (right).
		(b) Energy as a function of the in-plane magnetization direction $\phi_M^{ip}$ before (blue) and after (red) femtosecond laser excitation.
		The dashed black line sketches the in-plane deviation change during damping.
		(c) Magnetization precession trajectory in coordinates of in-plane vs. out-of-plane deviation.
		Open circles mark the energy minima.
		Black points show the period-averaged magnetization direction ($\phi_{\langle M\rangle}$).
	}
\end{figure}

To understand the physical mechanism behind the higher harmonics of the precession, we plot in Fig.~\ref{fig:energy_profile}\,(b) the energy profile (see Appendix~\ref{Sec:AppendixEnergy}) in the film plane as a function of the in-plane magnetization angle $\phi_M$ calculated under the laser excitation (red line) and for the equilibrium film (blue line).
The field of $\mu_0 H_{ext} = $30\,mT was chosen because it corresponds to the largest frequency shift in Fig.~\ref{fig:high_harmonics}\,(c).
Calculations confirm that the potential energy in the film plane becomes asymmetric under laser excitation.
Here, the potential energy is calculated within the macrospin approximation using the free energy density, with the magnetization constrained to the film plane ($\theta^{oop}_M = \pi/2$).
Precession in such a non-parabolic potential leads to both odd and even harmonics with the same polarization as the fundamental mode.

In the absence of damping, the precession follows an isoenergetic trajectory because the LLG equation contains no inertial terms. 
This yields a highly non-elliptical magnetization trajectory, shown in Fig.~\ref{fig:energy_profile}\,(c) in terms of out-of-plane $\phi_M^{oop}$ and in-plane  $\phi_M^{ip}$ angles.
Notably, the trajectory is mirror-symmetric only in the out-of-plane direction, indicating that the strong shape anisotropy creates a parabolic energy profile normal to the film plane and gives rise to strong ellipticity.
Aligning the major axis of the precession trajectory with the direction of potential asymmetry can significantly enhance the nonlinearity.

We also observe magnetic rectification, which manifests as a discrepancy between the period-averaged magnetization orientation $\phi_{\langle M\rangle}$ and the energy minimum after laser excitation $\phi^{min}_{M}$ [solid black and open red dots in Fig.~\ref{fig:energy_profile}, respectively].
This resembles the thermal expansion of solids~\cite{Dugdale1953}.
When an external field $\mathbf{H}_{ext}$ of strength comparable to the anisotropy field is applied close to HA, it reduces the barrier between the two equivalent anisotropy energy minima at $\pm \phi_M^{min}$.
A small in-plane deviation of $\mathbf{H}_{ext}$ from the HA lifts the degeneracy of the two minima~[Fig.~\ref{fig:energy_profile}\,(a)].
Rectification shifts $\phi_{\langle M\rangle}$ toward the flatter slope of the potential, i.e., toward the hard axis.
This reduces the precession frequency compared to the linear theoretical predictions [Fig.~\ref{fig:high_harmonics}\,(c)].

\begin{figure}
	\includegraphics[width= 0.9\linewidth]{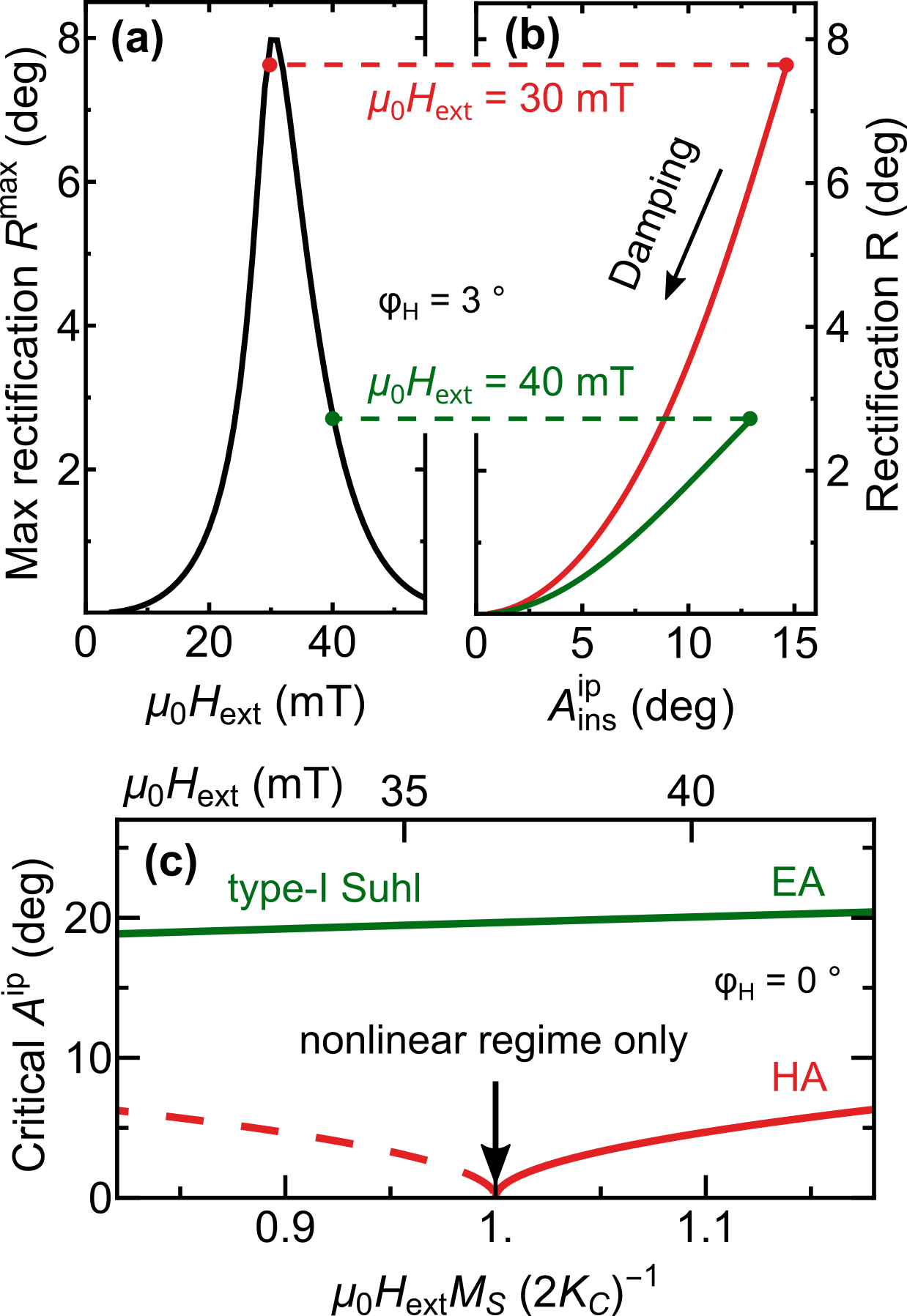}
	\caption{\label{fig:rectification_and_nonlinearity}
		{\bf Dependence of nonlinearity on the external magnetic field.}
		(a) Rectification $R = \phi_{\langle M\rangle} - \phi^{min}_{M}$ as a function of the external magnetic field at zero Gilbert damping.
		(b) Dependence of $R$ on the precession amplitude during damping for $\mu_0 H_{ext} = 30$ (red) and 40~mT (green).
		(c) Critical in-plane amplitude of nonlinearity vs. the ratio of Zeeman energy to anisotropy, given by Eq.~(\ref{eq:crit_angles}).
		Magnetization along EA (green) corresponds to type-I Suhl nonlinearity, while magnetization along HA (red) corresponds to anisotropy-induced nonlinearity.
		The arrow marks the field value at which a linear regime of dynamics is not realized.
	}
\end{figure}

To discuss magnetization precession with non-zero damping, we introduce the time-dependent rectification $R(\Delta t) = \phi_{\langle M\rangle}(\Delta t) - \phi^{min}_{M}$ and the instantaneous in-plane amplitude $A^{ip}_{ins}(\Delta t)$ as half the peak-to-peak value over a precession period~[Fig.~\ref{fig:energy_profile}\,(c)].
In this case, the magnetization follows a spiral trajectory towards equilibrium [black dashed lines in Fig.~\ref{fig:energy_profile}\,(b)]. 
Multiple precession cycles can be approximated by zero‑damping periods with varying amplitudes.
For fixed $\phi_H$, material parameters, and pump fluence, the maximum rectification $R^{max}$ immediately after excitation can be expressed as a function of $\mathbf{H}_{ext}$ [Fig.~\ref{fig:rectification_and_nonlinearity}\,(a)]. 
For finite damping, $R(\Delta t)$ decreases with the precession amplitude $A^{ip}_{ins}$ but persists until the dynamics is fully attenuated, indicating the absence of a threshold [see Fig.~\ref{fig:rectification_and_nonlinearity}\,(b)].

To substantiate the thresholdless character, we derive a critical in‑plane amplitude for the onset of nonlinearity (see Appendix~\ref{Sec:Critical_amplitude} for details).
The critical in-plane amplitudes are given by
\begin{equation}\label{eq:crit_angles}
	\begin{aligned}
		\Delta \phi_C^{\mathrm{HA}} &= \sqrt{\frac{c-1}{c-16}} \qquad \text{(HA)}, \\
		\Delta \phi_C^{\mathrm{EA}} &= \sqrt{\frac{c+1}{c+16}} \qquad \text{(EA, type-I Suhl)},
	\end{aligned}
\end{equation}
where $c = \mu_0 H_{\mathrm{ext}} M_S (2K_C)^{-1}$ is the ratio of the external field to the anisotropy field.
The field dependence of the critical amplitude for both cases, given by Eq.~(\ref{eq:crit_angles}), is shown in Fig.~\ref{fig:rectification_and_nonlinearity}\,(c).
In the EA case, $\Delta \phi_C^{\mathrm{EA}}$ is about $20$\,deg and varies only weakly with the field, indicating the presence of threshold for the Suhl nonlinearity.
In contrast, for the HA case $\Delta \phi_C^{\mathrm{HA}}$ vanishes as $H_{\mathrm{ext}}$ approaches the anisotropy field, revealing ultra‑small amplitude nonlinearity.
Note that for $H_{\mathrm{ext}}$ below the anisotropy field, the equilibrium magnetization is no longer collinear with the external field [dashed line in Fig.~\ref{fig:rectification_and_nonlinearity}\,(c)], and the above expression for $\Delta \phi_C^{\mathrm{HA}}$ is not applicable.

\subsection{\label{sec:waves} Spin waves}

\begin{figure*}
	\includegraphics[width= 0.9\linewidth]{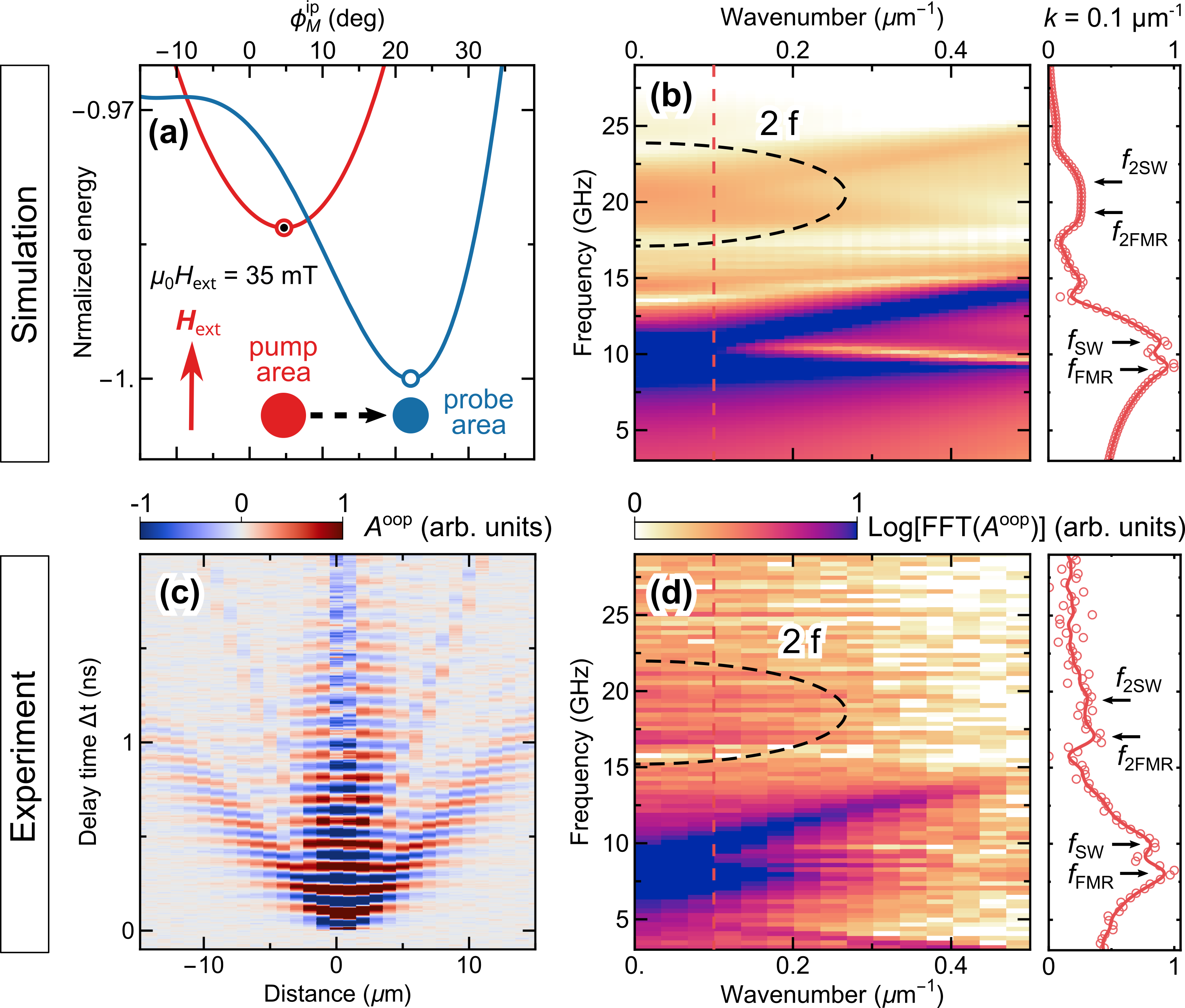}
	\caption{\label{fig:waves}
		{\bf Second harmonic of a propagating magnetostatic wave.}
		(a) Energy vs. in-plane magnetization direction inside (red) and outside (blue) the excitation area at $\mu_0 H_{ext} = 35$\,mT.
		(c) Map of the pump-probe signal vs. distance and delay time $\Delta t$).
		Measurements were performed at $\mu_0 H_{ext} = 35$\,mT directed at $\phi_H = 3$\,$^\circ$ in the Damon-Eshbach geometry.
		(b,d) 2D FFT of the micromagnetic simulation (b) and the experimental data (d) under the same conditions.
		The dashed black lines show the second harmonic of the magnetization dynamics.
		The right panels show cross-sections of the corresponding maps at $k = 0.1$~$\mu$m$^{-1}$ with a guide to the eyes (solid line).
		The cross-sections are indicated by red dashed lines on the maps.  
		The arrows on the cross-sections point to the spectral peaks of the first and second harmonics of the FMR and SSW.
	}
\end{figure*}

The pronounced nonlinearity allows us to observe the effect even at low amplitudes, including for propagating magnetostatic waves.
Indeed, let's consider the potential inside and outside the excitation region [red and blue lines in Fig.~\ref{fig:waves}\,(a), respectively]. 
After excitation, a wave packet propagates into regions of the film with equilibrium parameters.
If the potential is asymmetric in the propagation region, the wave should exhibit nonlinear properties.

We test this by generating the second harmonic of a magnetostatic surface spin wave (SSW) packet propagating perpendicular to $\mathbf{H}_{ext}$~[see inset in Fig.~\ref{fig:waves}\,(a)].
First, we perform a micromagnetic simulation~(Sec.~III of the Suppl. Material~\cite{Supplemental}).
The laser-induced reductions of $M_S$ and $K_C$ are 25\,\% and 94\,\%, respectively, corresponding to a mean laser fluence of $F = 39$\,mJ$\cdot$cm$^{-2}$. 
To create an asymmetric potential in the unexcited region at $\phi_H = 3$\,$^\circ$, the field is set to $\mu_0 H_{ext} = 35$\,mT [Fig.~\ref{fig:waves}\,(a)].
This geometry yields the longest propagation distance for SSW packets~\cite{Khokhlov_PRAppl2019}.
From the calculated spatial-temporal maps~(see details in Sec.~III Suppl. Material~\cite{Supplemental}), we obtain the SSW dispersion (in the nonlinear regime this dispersion is amplitude‑dependent).
The dispersion in Fig.~\ref{fig:waves}\,(b) shows a propagating SSW together with the wave at twice the frequency (encompassed by the black dashed line).
At $k = 0.1$ $\mu$m$^{-1}$, peaks corresponding to the FMR, SSW, and their doubled frequencies are visible [right panel of Fig.~\ref{fig:waves}\,(b)].
The second harmonic propagates at the same speed as the fundamental SSW packet (see the slope of the lines).
This confirms anharmonicity as the nonlinear mechanism -- the first and second harmonics are inseparable in space and time.
 
The same conditions were reproduced in the experiment.
The spatial-temporal map of the magneto-optical response shows propagating SSW packets [Fig.~\ref{fig:waves}\,(c)].
A 2D Fourier transform of the data reveals the wave dispersion [Fig.~\ref{fig:waves}\,(d)]. 
The weak signal below 5\,GHz in Fig.~\ref{fig:waves}\,(d) is a 2D-FFT artefact arising from the surface acoustic wave in the MgO substrate; it does not affect the spin-wave dynamics.
The second harmonic of the SSW packet was also observed.
However, the signal-to-noise ratio is insufficient to confirm its propagation via group velocity, i.e., the dispersion slope.
Nevertheless, at $k = 0.1$ $\mu$m$^{-1}$ a peak at the doubled frequency of the SSW is visible in the cross-section [right panel of Fig.~\ref{fig:waves}\,(d)], in good agreement with the simulation.
Subtracting the signal from the central region confirms the double-frequency signal beyond the excitation area~(see details in Sec.~IV Suppl. Material~\cite{Supplemental}).
This indicates that the second harmonic propagates at the same group velocity as the SSW.
The SSW also carries a rectified in-plane component, since the local magnetic moment trajectory resembles that in Fig.~\ref{fig:energy_profile}\,(c).
The presence of rectification over the entire amplitude range [Fig.~\ref{fig:rectification_and_nonlinearity}\,(b,c)] allows such waves to interact nonlinearly~\cite{breitbach2024nonlinear}.
As the wave propagates in the unexcited film, our conclusions apply to other excitation methods (e.g., by microwave antennas).

\section{\label{sec:level1} Conclusion}

In summary, we have experimentally demonstrated an intrinsic nonlinear mechanism in magnetization dynamics.
When an external magnetic field is applied close to the hard axis with a magnitude near the anisotropy field, the energy profile around the global minimum becomes asymmetric due to nearby local minima.
As a result, the dynamics are intrinsically nonlinear, and the Landau-Lifshitz-Gilbert equation cannot be linearized even for small amplitudes.
The magnetization dynamics exhibit three effects:
 (i) anharmonicity of the magnetization precession, which deviates from a single-frequency damped harmonic function;
 (ii) generation of both odd and even higher harmonics of the magnetization dynamics;
 and (iii) qualitative deviation of the fundamental precession frequency from the linear theory prediction.
The last effect is explained by magnetic rectification, which causes magnetization to oscillate around a point distinct from the energy minimum.

The presence of the effect at ultra‑small amplitude of magnetization dynamics enables excitation of nonlinear propagating spin-wave packets without requiring high power.
Our results establish a direct link between the magnetic energy landscape and nonlinear responses, providing a pathway for designing magnonic and spintronic devices that rely on controlled harmonic generation and spin-wave interactions.
The demonstrated nonlinearities only require a properly tailored magnetic anisotropy in combination with an external magnetic field or, more generally, other fields with the same symmetry, e.g. an exchange bias field.
Although we used laser-induced ultrafast heating to trigger the dynamics, the mechanism is not restricted to an excitation method; one can envisage exploiting various ultrafast opto-magnetic effects~\cite{kimel2020fundamentals} or even nonoptical approaches~\cite{barman20212021,flebus20242024}.
In ultrafast magnetism, our results have direct implications because laser-driven tuning of the precession frequency is an important goal~\cite{Wiechert2026,barman20212021,Shelukhin2022}, and also serves as a probe of the magnetic state following THz, infrared, and optical excitations~\cite{carpene2010ultrafast,schlauderer2019temporal,afanasiev2021ultrafast,soumah2021optical,Shelukhin2022,gerevenkov2021effect,kuzikova2025switching,schonfeld2025dynamical}.

\begin{acknowledgments}
The work of P.I.G. on pump-probe experiments, theoretical analysis, and micromagnetic simulations was supported by the grant of the RSF No. 24-72-00136, https://rscf.ru/project/24-72-00136/.
\end{acknowledgments}

\appendix
\section{\label{Sec:AppendixEnergy} Magnetic free energy}

The magnetic free energy density $E$ of the Fe(001) film includes the Zeeman term, the dipolar term, the cubic magnetocrystalline anisotropy, and the strain-induced easy-plane anisotropy:
\begin{equation}\label{eq:E}
\begin{aligned}
E = &-\mu_0 H_{ext} M_S \cos(\phi_M - \phi_H) \sin(\theta_M) \\
&+ \frac{1}{2} \mu_0 M_S^2 \cos(\theta_M) \\
&+ K_C \sin^2(\theta_M) [ \cos^2(\theta_M) +\\ 
&\sin^2(\theta_M) \cos^2\left(\phi_M - \frac{\pi}{4}\right) \sin^2\left(\phi_M - \frac{\pi}{4}\right) ] \\
&+ K_U \cos^2(\theta_M),
\end{aligned}
\end{equation}
where $\mu_0 = 4 \pi 10^{-7}$\,H~m$^{-1}$ is the vacuum permeability, $H_{ext}$ is the external in-plane magnetic field applied at an angle $\phi_H$ relative to the hard axis (HA), $M_S$ is the saturation magnetization, $K_{C}$ is the cubic anisotropy constant, and $K_U$ is the easy-plane anisotropy constant originating from interfacial strain. 
The angles $\phi_M$ and $\theta_M$ are the in-plane (relative to the HA) and out-of-plane (relative to the film normal) magnetization angles, respectively, as defined in Fig.~\ref{fig:high_harmonics}\,(b).
The equilibrium values of $\phi_{M0}$ and $\theta_{M0}$ are obtained numerically by minimizing $E$.

\section{\label{Sec:Critical_amplitude} Determination of the critical nonlinearity angle}

The classical type-I Suhl nonlinearity manifests itself as the dependence of the precession frequency on its amplitude.
This effect arises from the amplitude dependence of the period-averaged projection of the magnetization onto the direction of the effective field.
Consequently, the contribution of demagnetizing fields along the effective field decreases, while the transverse contribution increases.
For a small isotropic ellipsoid of revolution with the demagnetization tensor
\begin{equation}\label{eq:demag}
\mathbf{N} = \begin{pmatrix}
N_\perp & 0 & 0 \\
0 & N_\perp & 0 \\
0 & 0 & N_{||}
\end{pmatrix}
\end{equation}
and with the external magnetic field $\mathbf{H}_{ext}$ applied along the $N_{||}$ axis, the frequency-amplitude relation reads~\cite{gurevich1996magnetization}:
\begin{equation}\label{eq:nonlinearF}
\omega = -\gamma \left( \mu_0 H_{ext} - \mu_0 M_S \cos(\theta)(N_\perp - N_{||}) \right)
\end{equation}
where $\theta$ is the precession amplitude around the direction of the external magnetic field.

For an infinite crystal with cubic magnetic anisotropy $E_C = - \frac{1}{2} K_C \left( \left( \frac{M_x}{M_S} \right)^4 + \left( \frac{M_y}{M_S} \right)^4 + \left( \frac{M_z}{M_S} \right)^4 \right)$, one can introduce an effective diagonal demagnetizing tensor whose diagonal elements are given by $\frac{\partial E}{\partial M_i}$, where $M_i$ are the Cartesian components of the magnetization.
This representation allows extending Suhl's conclusions to the case of magnetization dynamics along the easy axis (EA) of the studied anisotropic film.

We now estimate the critical angle at which deviations from the linear regime are expected.
For this purpose, we use LLG without damping (Sec.~II B of Suppl. Material~\cite{Supplemental}).
Since the experimental signal is proportional to $\cos(\theta_M)$, we consider only the first equation of the system.
To separate linear and nonlinear contributions, we expand $E$ around the equilibrium position in powers of the small deviations $\Delta \phi_M$ and $\Delta \theta_M$ up to fourth order.
Taking into account that $\left. \frac{\partial E}{\partial \phi_M} \right|_{\phi_M=\phi_{M0},\; \theta_M=\theta_{M0}} = 0$ and differentiating with respect to $\Delta \phi_M$, and assuming $\Delta \theta \ll \Delta \phi$ (strong ellipticity of the film due to large $M_S$), we obtain:
\begin{equation}\label{eq:nonlinearE1}
\begin{aligned}
\frac{\partial E}{\partial \Delta \phi_M} &= \Delta \phi_M \left. \frac{\partial^{2} E}{\partial \phi_M^2} \right|_{\phi_M=\phi_{M0},\; \theta_M=\theta_{M0}}\\ &+ \frac{1}{2} \Delta \phi_M^2 \left. \frac{\partial^{3} E}{\partial \phi_M^3} \right|_{\phi_M=\phi_{M0},\; \theta_M=\theta_{M0}}\\ &+ \frac{1}{6} \Delta \phi_M^3 \left. \frac{\partial^{4} E}{\partial \phi_M^4} \right|_{\phi_M=\phi_{M0},\; \theta_M=\theta_{M0}}
\end{aligned}
\end{equation}
The first term in Eq.~(\ref{eq:nonlinearE1}) corresponds to linear in-plane dynamics, while the second and third terms represent nonlinear contributions.
Substituting the explicit form of the free energy Eq.~(\ref{eq:E}) into Eq.~(\ref{eq:nonlinearE1}), we get:
\begin{equation}\label{eq:nonlinearE2}
\begin{aligned}
\frac{\partial E}{\partial \Delta \phi_M} = &[ \mu_0 M_S H_{ext} \cos(\phi_{M0} - \phi_H) \sin(\theta_{M0})\\ &+ 2 K_C \cos\left(2 \phi_{M0} - \frac{\pi}{2}\right) \sin^4(\theta_{M0}) ] \Delta \phi_M\\ \\
&-\sin(\theta_{M0}) [ \mu_0 M_S H_{ext} \sin(\phi_{M0} - \phi_H)\\ &+ 8 K_C \sin^3(\theta_{M0}) \sin\left(4 \phi_{M0} - \pi\right) ] \Delta \phi_M^2\\ \\
&-\sin(\theta_{M0}) [ \mu_0 M_S H_{ext} \cos(\phi_{M0} - \phi_H)\\ &+ 32 K_C \sin^3(\theta_{M0}) \cos(4 \phi_{M0} - \pi) ] \Delta \phi_M^3
\end{aligned}
\end{equation}
As $\phi_H = \phi_{M0} = \frac{\pi}{4}$, $\theta_0 = \pi/2$, the term proportional to $\Delta \phi^2$ vanishes along high-symmetry directions but becomes allowed when deviating from them, as demonstrated in Ref.~\cite{marsh2012nonlinearly}.
The dynamics can be considered linear if the corresponding nonlinear term is negligible compared to the linear one.
We introduce the critical angle for the type-I Suhl nonlinearity as:
\begin{equation}\label{eq:crit_angle_EA}
\begin{gathered}
\left(\frac{\mu_0 M_S H_{ext}}{2 K_C} + 1 \right) \Delta \phi_C = \left( \frac{\mu_0 M_S H_{ext}}{2 K_C} + 16 \right) \Delta \phi_C^3 \\
\Delta \phi_C^{EA} = \sqrt{\frac{\frac{\mu_0 M_S H_{ext}}{2 K_C} + 1}{\frac{\mu_0 M_S H_{ext}}{2 K_C} + 16}}
\end{gathered}
\end{equation}

Now note that the linear part in Eq.~(\ref{eq:nonlinearE2}) is proportional to the effective field.
If we consider magnetization along the hard axis (HA) $\phi_H = \phi_{M0} = 0$ with an external field equal to the anisotropy field $\frac{2 K_C}{\mu_0 M_S}$, the linear part vanishes.
In that case, the nonlinear part exceeds the linear part for arbitrarily small amplitude.
Note that at the exact point where the linear part is zero, the dynamics vanishes because $\omega = 0$.
Nevertheless, in the vicinity of this point, the critical angle is less than one degree:
\begin{equation}\label{eq:crit_angle_HA}
\Delta \phi_C^{HA} = \sqrt{\frac{\frac{\mu_0 M_S H_{ext}}{2 K_C} - 1}{\frac{\mu_0 M_S 	H_{ext}}{2 K_C} - 16}}
\end{equation}
In this regime, the frequency shift occurs due to the rectification effect -- the deviation of $\langle \mathbf{M}\rangle_T$ from the equilibrium position.

\normalem
\bibliography{apssamp}

\end{document}